\documentclass[preprint,noshowpacs,preprintnumbers,amsmath,amssymb,superscriptaddress,nofootinbib]{revtex4}
\usepackage{graphicx,color}
\usepackage{amsmath,amssymb}
\usepackage{url}
\usepackage{epstopdf}
\usepackage{epsfig,graphicx}

\newcommand{\hs}{\hspace*{0.3cm}}

\newcommand{\eq}[1]{Eq.~(\ref{#1})}
\newcommand{\bib}[1]{Ref.~\cite{#1}}

\newcommand{\be}{\begin{equation}}
\newcommand{\ee}{\end{equation}}
\newcommand{\bea}{\begin{eqnarray}}
\newcommand{\eea}{\end{eqnarray}}

\newcommand{\crn}{\nonumber \\}

\newcommand{\fr}{\frac}
\newcommand{\bc}{\begin{center}}
\newcommand{\ec}{\end{center}}

\newcommand {\ba}{\begin{array}}
\newcommand {\ea}{\end{array}}
\newcommand{\ben}{\begin{enumerate}}
\newcommand{\een}{\end{enumerate}}

\begin{document}

\preprint{HU-EP-15/43}

\title{The simplest 3-3-1 model}
\author{Le Tho Hue}\email{lthue@iop.vast.ac.vn}
\affiliation{Institute of Physics,   Vietnam Academy of Science and Technology, 10 Dao Tan, Ba
Dinh, Hanoi, Vietnam }
\author{Le Duc Ninh}\email{Duc.Ninh.Le@physik.hu-berlin.de}
\affiliation{Humboldt-Universit{\"a}t zu Berlin, Institut f{\"u}r Physik, Newtonstrasse 15, 12489 Berlin, Germany}

\begin{abstract}
A simple extension of the Standard Model (SM), based on the gauge group $SU(3)_C\otimes SU(3)_L\otimes U(1)_Y$ with 
$Y$ being the hypercharge, is considered. 
We show that, by imposing an approximate global $SU(2)_L\times SU(2)_R$ custodial symmetry at the SM energy scale, the 
$Z-Z'$ mixing is absent at tree level and the value of the $\rho$ parameter can be kept close to one. 
Tree-level flavor-changing neutral currents are also reduced to three particles, namely $Z'$, a CP-odd Higgs and a CP-even Higgs. 
The model predicts new leptons with electric charges of $\pm 1/2e$ and new quarks with $\pm 1/6e$ charges 
as well as new gauge and scalar bosons with $\pm 1/2e$ charges. 
Electric charge conservation requires that one of them must be stable. 
Their masses are unfortunately free parameters.

\end{abstract}
\maketitle
\section{Introduction}
\label{sec:intro}
Simple extensions of the Standard Model (SM), based on the gauge group $SU(3)_C\otimes SU(3)_L\otimes U(1)_X$, have been extensively studied (see \bib{Singer:1980sw} 
and references therein, see also Refs.~\cite{Valle:1983dk,Pisano:1991ee,Foot:1992rh,Frampton:1992wt,Foot:1994ym} for similar models 
but with lepton-number violation). There are many 3-3-1 models different mainly at the choice of fermion content and representations. 
Typically fermions are organized into triplets and anti-triplets of $SU(3)_L$ in three generations. It is also usually required that the 
SM is recovered at low energies. With the present data \cite{Agashe:2014kda} there seems to be still a lot of freedom in choosing the 
third entries of the (anti-)triplets. For example, one can put in new heavy leptons or the anti-particles of the known leptons. 
One can also choose to have new electric charges, e.g. new quarks with $Q=\pm 4/3$ and $\pm 5/3$ as in Refs.~\cite{Pisano:1991ee,Frampton:1992wt}. 

It is convenient to classify 3-3-1 models using the $\beta$ parameter defined via the electric charge operator
\bea
Q = \alpha T_3 + \beta T_8 + X,
\label{eq:charge_Q}
\eea
where we have introduced the $SU(3)$ generators $T_a$, $a=1,\ldots 8$ and $X$ the new quantum charge corresponding to 
the group $U(1)_X$. To match with the SM, where $Q_\text{SM} = I_3 + Y/2$ with $I_3$ being the weak isospin generator and 
$Y$ the weak hypercharge, we must have $\alpha = 1$ and 
\bea
\fr{Y}{2} = \beta T_8 + X.
\label{eq:hypercharge}
\eea
When the fermions and their representations are fixed then the value of $\beta$ is uniquely determined. The reverse is however not 
true. Knowing $\beta$ fixes the electric charges (for given representations) but not other properties such as lepton/baryon number or mass. 

Most studies so far have focused on the case of $\beta = \pm n/\sqrt{3}$ with $n=1$ or $3$. 
Studies with so-called arbitrary $\beta$ have also been done, see e.g. Refs.~\cite{Diaz:2004fs,Buras:2012dp}. 
Matching the couplings with the SM leads approximately to the constraint $n \leq 3$, see e.g. \bib{Diaz:2004fs}. 
If we require that the electric charges of the leptons 
and quarks must be the same as in the SM then $n=1$. All 3-3-1 models have a new distinct feature compared to the SM, 
namely flavor-changing neutral current (FCNC) effects occur at tree level. This happens in the gauge and Higgs sectors \cite{Montero:1992jk}. The new 
neutral gauge boson $Z'$ induces FCNC because anomaly cancellation forces one 
family of quarks to behave differently from the other quark families. FCNCs in the Higgs sector are also due to this reason but also due to the fact that 
there are more than one scalar triplets. Another important feature is $Z-Z'$ mixing. This causes FCNC at low energies and also introduces correction to 
$\rho$ parameter, defined as $\rho = m^2_W/(m^2_Z\cos^2\theta_W)$ with $\theta_W$ being the weak-mixing angle, at tree level. A popular solution is to 
suppress the mixing by requiring $m_{Z'} \gg m_{Z}$. 
However, since the mixing depends on the vacuum expectation values (VEV), we can also kill this mixing at 
tree level by imposing that the VEVs satisfy a certain condition \cite{VanDong:2005pi,Dias:2005xj}. It is therefore hoped that $m_{Z'}$ is not so far from the SM electroweak (EW) scale \cite{Dias:2005xj}. The $Z-Z'$ mixing also 
breaks the $\beta \leftrightarrow -\beta$ and simultaneously triplet $\leftrightarrow$ anti-triplet symmetry, see e.g. \bib{Buras:2014yna}.         
From a practical viewpoint, this mixing makes the Feynman rules complicated and hence the models less attractive. 

The above consideration leads us to an important remark: the simplest picture emerges in the case $\beta = 0$. 
We will show in this letter that, only in this case, the requirement of no $Z-Z'$ mixing at tree level (barring the decoupling limit) 
leads to a very simple constraint on the VEVs of the two scalar triplets responsible for the
symmetry breaking from $SU(2)_L\otimes U(1)_Y$ to $U(1)_Q$, namely $v' = v$. This is also a hint to obtain a simple form 
of the scalar potential by imposing an approximate custodial symmetry \cite{Sikivie:1980hm} at low energies. 
It follows that the value of the $\rho$ parameter can be kept close to one 
and FCNCs in the scalar sector are reduced, being restricted to one CP-odd and one CP-even Higgs bosons.  
To the best of our knowledge, the 3-3-1 model with $\beta = 0$ has never been considered in the literature 
\footnote{This case was excluded in Refs.~\cite{Diaz:2003dk,Diaz:2004fs} without justification. We thank 
R.~Martinez for discussion on this issue.}. 

\section{The model}
With $\beta = 0$, the model can be named $331Y$ based on the gauge group $SU(3)_C\otimes SU(3)_L\otimes U(1)_Y$, we can write down 
the fermion representation as follows. Left-handed leptons are assigned to anti-triplets and right-handed leptons are 
singlets: 
\bea && L_{aL}=\left(
       \begin{array}{c}
         e_a \\
         -\nu_{a} \\
         E_a \\
       \end{array}
     \right)_L \sim \left(3^*, -1\right), \hs a=1,2,3,\crn
     && e_{aR}\sim   \left(1, -2\right)  , \hs \nu_{aR}\sim  \left(1, 0\right) ,\hs E_{aR} \sim   \left(1, -1\right),  \label{lep}\eea
  where the introduction of three right-handed neutrino states is optional and the new leptons $E^a_{L,R}$ have electric charges equal to $-1/2$ (from now on we 
use the unit of the proton charge). 
  The numbers in the parentheses are to label the representation of $SU(3)_L\otimes U(1)_Y$ group. 

  From now on we leave the $SU(3)_C$ group aside, since it is the same as in the SM. 
  In order to cancel all triangle anomalies the number of anti-triplets must be equal to the number of triplets. 
  This means that one generation of quarks must be anti-triplets and two generations are triplets. 
  In other words, there are 6 anti-triplets of leptons and quarks (which come in 3 colors) and 6 triplets of quarks. 
  We choose that the first two generations of quarks are in triplets and the third anti-triplet as
\bea && Q_{iL}=\left(
       \begin{array}{c}
         u_i \\
         d_i \\
         U_i \\
       \end{array}
     \right)_L \sim \left(3, 1/3\right), \hs i=1,2,\crn
&& Q_{3L}=\left(     
       \begin{array}{c}
         b \\
         -t \\
         T \\
       \end{array}
     \right)_L \sim \left(3^*, 1/3\right), \crn
&& u_{iR}\sim   \left(1, 4/3\right)  , \hs d_{iR}\sim  \left(1, -2/3\right) ,\hs U_{iR} \sim \left(1, 1/3\right),\crn
&& t_{R}\sim   \left(1, 4/3\right)  , \hs b_{R}\sim  \left(1, -2/3\right) ,\hs T_{R} \sim \left(1, 1/3\right), 
\label{quark}\eea
where we have introduced three new quarks $U_1$, $U_2$ and $T$ with electric charges all equal to $1/6$. This is different from 
models with $\beta \neq 0$, where different charges are predicted as $Q_{U_i} = 1/6 - \sqrt{3}\beta/2$ and $Q_{T} = 1/6 + \sqrt{3}\beta/2$. 

Unlike the SM, where anomaly cancellation happens within one generation of leptons and quarks, 
the anomaly cancellation in 3-3-1 models occurs after summing over leptons and quarks of three generations. 
The key difference is that the representations of $SU(2)$ are real, while it is not the case for $SU(3)$. 
This is why we need an equal number of anti-triplets and triplets, which forces one family of quarks 
to behave differently from the other two families as above mentioned. 

We now discuss the gauge sector.  
There are totally nine EW gauge bosons, included in the following covariant derivative
 \bea D_{\mu}\equiv \partial_{\mu}-i g_3 T^a W^a_{\mu}-i g_1 \fr{Y}{2} B_{\mu},  \label{coderivative1}\eea
where $g_3$ and $g_1$ are coupling constants corresponding to the two groups $SU(3)_L$ and $U(1)_Y$, respectively. 
The matrix $W^aT^a$, with $T^a =\lambda_a/2$ corresponding to a triplet representation, can be written as
 \bea W^a_{\mu}T^a=\frac{1}{2}\left(
                     \begin{array}{ccc}
                       W^3_{\mu}+\frac{1}{\sqrt{3}} W^8_{\mu}& \sqrt{2}W^+_{\mu} &  \sqrt{2}V^{+1/2}_{\mu} \\
                        \sqrt{2}W^-_{\mu} &  -W^3_{\mu}+\frac{1}{\sqrt{3}} W^8_{\mu} & \sqrt{2}V'^{-1/2}_{\mu} \\
                       \sqrt{2}V^{-1/2}_{\mu}& \sqrt{2}V'^{+1/2}_{\mu} &-\frac{2}{\sqrt{3}} W^8_{\mu}\\
                     \end{array}
                   \right),
  \label{wata}\eea
where we have defined the mass eigenstates of the charged gauge bosons as
\bea W^{\pm}_{\mu}=\frac{1}{\sqrt{2}}\left( W^1_{\mu}\mp i W^2_{\mu}\right),\crn
V^{\pm 1/2}_{\mu}=\frac{1}{\sqrt{2}}\left( W^4_{\mu}\mp i W^5_{\mu}\right),\crn
V'^{\mp 1/2}_{\mu}=\frac{1}{\sqrt{2}}\left( W^6_{\mu}\mp i W^7_{\mu}\right).
   \label{gbos}\eea 
We notice that in addition to the SM gauge bosons there is one more neutral gauge boson and four new charged gauge bosons with $Q = \pm 1/2$.

To generate masses for gauge bosons and fermions, we need three scalar triplets. They are 
defined as
  \bea  
   \eta=\left(
              \begin{array}{c}
                \eta^0 \\
                \eta^- \\
                \eta^{-1/2} \\
              \end{array}
            \right)\sim \left(3, -1\right), \hs
    \rho=\left(
              \begin{array}{c}
                \rho^+ \\
                \rho^0 \\
                \rho^{+1/2} \\
              \end{array}
            \right)\sim \left(3, 1\right), \hs
    \chi=\left(
              \begin{array}{c}
                \chi^{+1/2} \\
                \chi'^{-1/2} \\
                \chi^0 \\
              \end{array}
            \right)\sim \left(3, 0\right).
    \label{higgsc}
  \eea
These Higgses develop VEVs as
  \bea 
\langle   \eta \rangle= \frac{1}{\sqrt{2}}\left(
              \begin{array}{c}
                v' \\
                0 \\
                0 \\
              \end{array}
            \right), \hs 
 \langle  \rho \rangle =\frac{1}{\sqrt{2}}\left(
              \begin{array}{c}
                0 \\
                v \\
                0 \\
              \end{array}
            \right),\hs
\langle  \chi\rangle=\frac{1}{\sqrt{2}}\left(
              \begin{array}{c}
                0 \\
                0 \\
                u \\
              \end{array}
            \right).
            \label{vevhigg1}\eea
 The symmetry breaking happens in two steps:  
\bea 
SU(3)_L\otimes U(1)_Y\xrightarrow{u} SU(2)_L\otimes U(1)_Y\xrightarrow{v,v'} U(1)_Q.
\label{symmetry_breaking}
\eea  
It is therefore reasonable to assume that $u > v,v'$. 
After the first step, 
five gauge bosons ($W^8$, $V^{\pm 1/2}$ and $V'^{\pm 1/2}$) will be massive and the remaining four massless gauge bosons 
can be identified with the before-symmetry-breaking SM gauge bosons. The new neutral gauge boson $W^8$ is already a physical state and is called $Z'$. After the second breaking, we obtain the following results: 
\bea
Z'_\mu = W^8_\mu, \hs \left(\begin{array}{c} Z_\mu \\ A_\mu \end{array}\right) = 
\left(\begin{array}{cc} c_W & -s_W\\ s_W & c_W \end{array}  \right)\left(\begin{array}{c} W^3_\mu \\ B_\mu \end{array}\right),
\eea
where $c_W = \cos\theta_W$, $s_W = \sin\theta_W$ with $\theta_W$ being the weak mixing angle read
\bea
s_W = \fr{g_1}{\sqrt{g_1^2 + g_3^2}}, \hs c_W = \fr{g_3}{\sqrt{g_1^2 + g_3^2}}.
\eea

The masses of the charged gauge bosons are
\bea 
m^2_{V} =\frac{g_3^2}{4}(u^2 + v'^2),\quad m^2_{V'} = \frac{g_3^2}{4}(u^2 + v^2), \quad m^2_{W} = \fr{g_3^2}{4}(v^2 + v^{\prime 2}).
\label{mass_gauge_charged}\eea
The mass matrix for the neutral gauge bosons ($A$, $Z$, $Z'$) reads
\bea
   \mathcal{M}^2 = \left(
                 \begin{array}{ccc}
                   0& 0& 0\\
                   0& M_{ZZ}^2 & M_{ZZ'}^2 \\
                   0& M_{ZZ'}^2 & M_{Z'Z'}^2 \\
                 \end{array}
               \right),
 \eea
where
\bea
M_{ZZ}^2 = \fr{g_3^2}{4c_W^2} (v^2 + v'^{2}), \hs M_{Z'Z'}^2 = \fr{g_3^2}{12}\left(4u^2 + v^2 + v'^2 \right)
\eea
and the off-diagonal entry is
\bea
M_{ZZ'}^2 = \fr{g_3^2}{4\sqrt{3}c_W}(v'^2 - v^2). 
\label{eq_M23}
\eea
This result shows us clearly that if we demand $v = v'$ then $Z$ and $Z'$ do not mix and 
the $\rho$ parameter is exactly one at tree level. We then obtain 
\bea
&& m_W^2 = \fr{g_3^2v^2}{2}, \hs
m_Z = \fr{m_W}{c_W}, \crn
&& m_V^2 = m_{V'}^2 = \fr{g_3^2}{4}(u^2 + v^2), \hs
m^2_{Z'} = \fr{g_3^2}{6}(2u^2 + v^2).
\label{mass_gauge_final}
\eea
At this point we note that the condition of no $Z-Z'$ mixing, in the scenario where the scale $u$ is not so far 
from $v,v'$, has been extensively discussed in \bib{Dias:2005xj,Dias:2006ns} for the cases of $\beta = -\sqrt{3}$ and $\beta = -1/\sqrt{3}$. 
The same condition is also noted in \bib{VanDong:2005pi} for the general case with arbitrary $\beta$, which reads
\bea
v^2 = \fr{1+(\sqrt{3}\beta - 1)s_W^2}{1 - (\sqrt{3}\beta + 1)s_W^2}v'^2.
\label{dong_u_v}
\eea
We see that, if $\beta = 0$, the above condition $v=v'$ is again obtained. 
If $\beta \neq 0$ then we have $v \neq v'$ and hence $m_V \neq m_{V'}$ (see \eq{mass_gauge_charged}). This means that, when one-loop corrections to the $\rho$ parameter 
are considered, there is 
a fundamental difference between the two cases due to the contribution of the new gauge bosons to the oblique parameter $T$ which is, in the absence of $Z-Z'$ mixing, proportional to the mass splitting $(m_{V} - m_{V'})$ as shown in \bib{Hoang:1999yv}. This correction is zero for $\beta = 0$ and non-zero otherwise. This suggests that the global custodial symmetry (see below) is broken if $\beta \neq 0$. 
 
We note, in passing, that \eq{mass_gauge_final} gives
\bea
\fr{m_{Z'}}{m_V} \approx \sqrt{\fr{4}{3}} \approx \fr{m_Z}{m_W},
\eea
if the condition $u \gg v$ (or $m_{Z'}\gg m_Z$) is assumed for the first approximation. 

We now turn to the scalar sector. The most general scalar potential, which is renormalizable 
and gauge invariant, reads
\bea
\mathcal{V} &=& \mu_1^2 \eta^\dagger \eta + \mu_2^2 \rho^\dagger \rho + \mu_3^2 \chi^\dagger \chi
+ \lambda_1 (\eta^\dagger \eta)^2 + \lambda_2 (\rho^\dagger \rho)^2 + \lambda_3 (\chi^\dagger \chi)^2\crn
&+& \lambda_{12} (\eta^\dagger \eta)(\rho^\dagger \rho)  + \lambda_{13} (\eta^\dagger \eta)(\chi^\dagger \chi) 
+ \lambda_{23} (\rho^\dagger \rho)(\chi^\dagger \chi)\crn 
&+& \tilde\lambda_{12} (\eta^\dagger \rho)(\rho^\dagger \eta)  + \tilde\lambda_{13} (\eta^\dagger \chi)(\chi^\dagger \eta) 
+ \tilde\lambda_{23} (\rho^\dagger \chi)(\chi^\dagger \rho)\crn 
&+& \sqrt{2}f (\epsilon_{ijk}\eta^i \rho^j \chi^k + h.c.), 
\label{potential}
\eea 
where $f$ is a mass parameter and is assumed to be real. 
This potential has been studied in Refs.~\cite{Nguyen:1998ui,Tonasse:1996cx}. Minimizing the potential with respect to $u$, $v$ and $v'$, we get
\begin{align}
\mu_1^2 + \lambda_1 v'^2 +\fr{1}{2}\lambda_{12} v^2 + \fr{1}{2}\lambda_{13} u^2 &= -f\fr{vu}{v'},\crn
\mu_2^2 + \lambda_2 v^2 + \fr{1}{2}\lambda_{12} v'^2+ \fr{1}{2}\lambda_{23} u^2 &= -f\fr{v'u}{v},\label{minimum_condition}\\
\mu_3^2 + \lambda_3 u^2 + \fr{1}{2}\lambda_{13} v'^2+ \fr{1}{2}\lambda_{23} v^2 &= -f\fr{vv'}{u}.\nonumber 
\end{align}
Requiring the potential to be bounded from below gives $\lambda_i > 0$ with $i=1,2,3$. 
The mixing pattern is the same as in the minimal model~\cite{Pisano:1991ee} and has been studied in \bib{Nguyen:1998ui}. 
One special point to notice is that, even though there are four scalar fields with the same electric charge ($Q=\pm 1/2$), 
they mix in pairs as in the minimal model, 
namely 
($\eta^{\pm 1/2}$, $\chi^{\pm 1/2}$), ($\rho^{\pm 1/2}$, $\chi'^{\pm 1/2}$), 
to form four charged Goldstone bosons denoted $G^{\pm 1/2}_V$ and $G^{\pm 1/2}_{V'}$ as well as 
four charged Higgses $H^{\pm 1/2}$ and $H'^{\pm 1/2}$. 
Similarly, the singly charged scalars $\eta^\pm$ and $\rho^\pm$ mix to form two $W$ Goldstone bosons $G^{\pm}_W$ and 
two charged Higgses $H^{\pm}$.  
The neutral scalars are defined as
\bea
\eta^0 = \fr{1}{\sqrt{2}}(v' + h_1 + i\zeta_1), \quad \rho^0 = \fr{1}{\sqrt{2}}(v + h_2 + i\zeta_2), \quad 
\chi^0 = \fr{1}{\sqrt{2}}(u + h_3 + i\zeta_3).
\eea
For the pseudoscalar bosons, the mass matrix in the 
basis ($\zeta_1$, $\zeta_2$, $\zeta_3$) reads: 
\bea
   \mathcal{M}_A^2 = -f u \left(
                 \begin{array}{ccc}
                   v/v'& 1& v/u\\
                   1& v'/v & v'/u \\
                   v/u& v'/u & vv'/u^2 \\
                 \end{array}
               \right).
 \eea
For $v = v'$, we then obtain two massless Goldstone bosons and one CP-odd Higgs named $A$ with mass:
\bea
m_A^2 = -fu\left(2 + t^2\right),\quad t = v/u.
\label{def_mA_t}
\eea
This result requires $f\le 0$. 
The rotation matrix reads:  
\bea
   \left(\begin{array}{c} \zeta_1\\ \zeta_2\\ \zeta_3 \end{array} \right) 
    = \left( \begin{array}{ccc}
               -\fr{1}{\sqrt{2}}& -\fr{t}{\sqrt{2(2+t^2)}}& \fr{1}{\sqrt{2+t^2}}\\
                   \fr{1}{\sqrt{2}}&    -\fr{t}{\sqrt{2(2+t^2)}}& \fr{1}{\sqrt{2+t^2}} \\
                   0&   \fr{\sqrt{2}}{\sqrt{2+t^2}}&  \fr{t}{\sqrt{2+t^2}} \\
                 \end{array} 
               \right)\left(\begin{array}{c} G_Z\\ G_{Z'}\\ A \end{array} \right).
\label{mixing_CP_odd_Higgs}
\eea
The three CP-even Higgses also mix to form the physical states. The mass matrix in the basis ($h_1$, $h_2$, $h_3$) is given by
\bea
   \mathcal{M}_H^2 = \left(
                 \begin{array}{ccc}
                   2\lambda_1 v'^2 - fvu/v' & \lambda_{12}vv' + fu & \lambda_{13}v'u + fv \\
                   \lambda_{12}vv' + fu & 2\lambda_2 v^2 - fv'u/v & \lambda_{23}vu + fv' \\
                   \lambda_{13}v'u + fv & \lambda_{23}vu + fv' & 2\lambda_3 u^2 - fvv'/u \\
                 \end{array}
               \right),\label{matrix_CP_even_Higgs}
 \eea
which agrees with \bib{Nguyen:1998ui}.
We observe that, even with the condition $v = v'$ there is no simple solution for the Higgs masses. 
This is one of the most difficult problems of 3-3-1 models, namely the scalar potential contains many parameters. 
It is therefore desirable to identify a simple form of the potential such that the CP-even Higgs masses and couplings to 
other particles can be easily calculated and the SM physics can be obtained at low energies. We found that $v = v'$ 
is a very important hint to achieve this. Indeed, from \eq{minimum_condition} we see that the equality can be obtained if 
\bea
\mu_1 = \mu_2, \hs \lambda_1 = \lambda_2, \hs \lambda_{13} = \lambda_{23}.
\label{custodial_cond_1}
\eea
Where does this come from? Is it related to any symmetry? 

If we impose that the approximate global custodial symmetry $SU(2)_L\times SU(2)_R$ of the SM is satisfied by the scalar potential, 
we will also get back the constraints in \eq{custodial_cond_1} plus other constraints. 
This can be seen as follows. The scalar fields involved in the global custodial symmetry at the SM energy scale are: 
 \bea  
   \eta'=\left(
              \begin{array}{c}
                \eta^0 \\
                \eta^- 
              \end{array}
            \right), \hs
    \rho'=\left(
              \begin{array}{c}
                \rho^+ \\
                \rho^0 
              \end{array}
            \right).
    \label{higgsc_2}
  \eea
We then define, as usual, $\Phi = (\eta'\; \rho')/\sqrt{2}$, and write down the most general potential symmetric under $SU(2)_L\times SU(2)_R$, which reads (see e.g. \bib{Pomarol:1993mu})
\bea
\mathcal{V}^\text{SM}_C &=& m_1^2 \text{Tr}(\Phi^+\Phi) + \left[m_2^2 \text{det}(\Phi) + h.c.\right] + \lambda [\text{Tr}(\Phi^+\Phi)]^2 + \lambda_4 \text{det}(\Phi^+\Phi)\crn
&+& \left[\lambda_5 (\text{det}\Phi)^2 + h.c.\right] + \left[\lambda_6 \text{det}\Phi \text{Tr}(\Phi^+\Phi) + h.c. \right].
\label{potential_C}
\eea
The symmetry is broken down to $SU(2)_V$ if $v=v'$.
By requiring that the potential in \eq{potential} matches \eq{potential_C} for any values of $\chi$, $\eta^{-1/2}$ and $\rho^{+1/2}$, and ignoring the terms linear in 
any component of $\eta'$ or $\rho'$, we get back the condition in \eq{custodial_cond_1} and\bea
\lambda_{12} = 2\lambda_1,\quad
\tilde{\lambda}_{12}  = \tilde{\lambda}_{13} = \tilde{\lambda}_{23} = 0.
\label{custodial_cond_2}
\eea
This means that the terms proportional to $\tilde{\lambda}_{12}$, $\tilde{\lambda}_{13}$, or $\tilde{\lambda}_{23}$ in the scalar potential break the custodial symmetry, and hence can 
give large corrections to the $\rho$ parameter in the general case where the custodial symmetry is not imposed. 

It is important to note that the conditions in \eq{custodial_cond_1} and \eq{custodial_cond_2} are {\em tree-level} relations 
and that they are broken by radiative corrections. This is because the custodial symmetry is broken by Yukawa and $U(1)_Y$ interactions. One-loop corrections to those relations are, in general, nonvanishing and divergent, as shown in \bib{Gunion:1990dt} for an 
extended version of the SM with two scalar triplets. To cancel all UV divergences we need enough counterterms. 
It is therefore important to keep in mind that, some parameters may be absent at tree level (due to the custodial symmetry) 
but their counterterms are needed at one-loop level to obtain finite results. In other words, the conditions in \eq{custodial_cond_1} and \eq{custodial_cond_2} cannot be imposed on the counterterms. 

Since the custodial symmetry is just an approximate symmetry, the above conditions should be also approximate. They 
should be understood as guidelines for keeping $\rho$ close to one. Radiative corrections break those conditions and 
hence induce {\em corrections} to $\rho$. These corrections are higher-order effects and should be, in principle, 
small. However, some quadratic divergences can occur, as shown in \bib{Gunion:1990dt}, and hence one-loop corrections may be significant. Nevertheless, these large corrections can be cancelled by fine-tuning the counterterm of $\rho$ \cite{Gunion:1990dt}. 
Our conclusion is therefore that the value of $\rho$ can be kept close to one in the present model with the 
relations in \eq{custodial_cond_1} and \eq{custodial_cond_2} approximately satisfied.      

We now impose the conditions in \eq{custodial_cond_1} and \eq{custodial_cond_2} to find the eigenvalues and eigenvectors of the matrix in \eq{matrix_CP_even_Higgs}. 
It turns out that a simple solution exists. The mass matrix can now be written as
\bea
\mathcal{M}^2_H = u^2 \tilde{\mathcal{M}}^2_H,\quad 
\tilde{\mathcal{M}}^2_H = 
               \left(
                 \begin{array}{ccc}
                   a & b & c \\
                   b & a & c \\
                   c & c & d \\
                 \end{array}
               \right),\label{matrix_CP_even_Higgs_simple}
\eea
where
\bea
a = 2\lambda_1 t^2 - f/u,\hs b = 2\lambda_1 t^2 + f/u, \hs c = (\lambda_{13} + f/u)t, \hs d = 2\lambda_3 - t^2 f/u. 
\label{def_abcd}
\eea 
The three CP-even Higgses have the following masses~\footnote{The diagonalization can be done in two symmetry-breaking steps with $v=0$ in the first step.}
\bea
m^2_{H_1} &=& 
\frac{1}{2} u^2 \left(a + b + d - \sqrt{\Delta} \right),\hs m^2_{H_2} = 
\frac{1}{2} u^2 \left(a + b + d + \sqrt{\Delta} \right),\crn
\Delta &=& (a+b-d)^2 + 8c^2,\hs
m^2_{H_3} = u^2(a-b) = -2 f u.
\label{mass_CP_even}
\eea
We define the 
physical CP-even Higgs bosons via 
\bea
\left(\begin{array}{c} h_1\\ h_2\\ h_3 \end{array}\right) = 
               \left(
                 \begin{array}{ccc}
                   -\fr{c_\alpha}{\sqrt{2}} & \fr{s_\alpha}{\sqrt{2}} & -\fr{1}{\sqrt{2}} \\
                   -\fr{c_\alpha}{\sqrt{2}} & \fr{s_\alpha}{\sqrt{2}} & \fr{1}{\sqrt{2}} \\
                   s_\alpha & c_\alpha & 0 \\
                 \end{array}
               \right)\left(\begin{array}{c} H_1\\ H_2\\ H_3 \end{array}\right), 
\label{mixing_CP_even_Higgs}
\eea
where $s_\alpha = \sin\alpha$ and $c_\alpha = \cos\alpha$ and they are defined by
\bea
s_\alpha = \fr{a + b - \kappa_1}{\sqrt{2c^2 + (a+b-\kappa_1)^2}},\hs 
c_\alpha = \fr{\sqrt{2}c}{\sqrt{2c^2 + (a+b-\kappa_1)^2}}, 
\hs \kappa_1 = m_{H_1}^2/u^2.
\label{angle_alpha}
\eea
In the decoupling limit $t\ll 1$ we get, (recall $\lambda_3 > 0$),
\bea
m_{H_1}^2 \approx v^2\left[4\lambda_1 - \fr{(\lambda_{13}+f/u)^2}{\lambda_3} \right],\hs
m_{H_2}^2 \approx 2\lambda_3 u^2,\hs s_\alpha \approx \fr{(\lambda_{13}+f/u)t}{\sqrt{2}\lambda_3}.\label{mH_approx}
\eea
We observe that the two Higgses $H_2$ and $H_3$ get masses after the first symmetry breaking and 
$H_1$ gets mass after the second breaking. $s_\alpha$ is suppressed, meaning that $H_1$ couples very weakly to the 
new fermions. Therefore, $H_1$ is similar to the SM Higgs. 

We are now in the position to examine the FCNCs in the scalar sector. 
For this purpose, we need to consider Yukawa interactions. 
For the leptons, since the three families transform identically under the $SU(3)_L$ group, there is no FCNC because 
diagonalizing the mass matrices automatically makes the interactions diagonal. For the quark sector, it is more complicated because the 
third family transforms differently. The Lagrangian reads: 
\bea
\mathcal{L}_\text{yuk} &=& -Y^u_{ia} \bar{Q}_{iL} \eta u_{aR} - Y^d_{ia} \bar{Q}_{iL} \rho d_{aR} - Y^U_{ia} \bar{Q}_{iL} \chi U_{aR}\crn
&-& Y^d_{3a} \bar{Q}_{3L} \eta^* d_{aR} - Y^u_{3a} \bar{Q}_{3L} \rho^* u_{aR} - Y^U_{3a} \bar{Q}_{3L} \chi^* U_{aR} + h.c., 
\eea 
where $i = 1,2$; $a = 1,2,3$; $u_{aR} = u_R, c_R, t_R$; $d_{aR} = d_R, s_R, b_R$ and $U_{aR} = U_{1R}, U_{2R}, T_R$. 
From this, together with \eq{mixing_CP_odd_Higgs} and \eq{mixing_CP_even_Higgs}, we can easily see (by examining the mass matrices 
and interaction matrices) that 
there is no FCNC related to the Goldstone boson of the $Z$ as expected. 
More interesting is that, thanks to the custodial symmetry, the two CP-even Higgses $H_1$ and $H_2$ do not induce FCNC, 
which is in general not the case if $v \neq v'$. One of these Higgses can be identified with the SM Higgs. 
For example, it is $H_1$ in the decoupling limit. The other neutral Higgs bosons, namely the CP-odd Higgs and $H_3$, do induce FCNCs. 
It may be a mistake, therefore, to conclude that constraints on FCNC implies that the $Z'$ is very heavy 
because destructive interference effects can occur, as discussed in Refs.~\cite{Dias:2005xj,Machado:2013jca}. 

We note that the conditions in \eq{custodial_cond_1} and \eq{custodial_cond_2} from the approximate custodial symmetry at low energies 
can also be applied to any model with the same scalar potential. 
However, differently to the present model, the tree-level $Z-Z'$ mixing remains (proportionally to $\beta$).  

Finally we have a few comments on the exotic-charged particles, 
namely three exotic quarks with electric charges of $\pm 1/6e$, three 
exotic leptons with charges $\pm 1/2e$ as well as new gauge and scalar bosons with charges $\pm 1/2e$. 
Electric charge conservation requires that one of them must be stable, being either a fermion or a boson. 
If it is a fermion, say the lepton $E_1$, then we can have the following signature at the LHC. 
A pair of exotic quarks can be produced via gluon fusion, followed by subsequent decays to the stable lepton: 
\bea
&&E^{-1/2}_2 \to \mu^- V^{* +1/2} \to \mu^- e^+ E^{-1/2}_1,\crn
&&T^{+1/6} \to b^{-1/3} V^{* +1/2} \to b^{-1/3} e^+ E^{-1/2}_1,\crn
&&U^{+1/6}_1 \to d^{-1/3} V'^{* +1/2} \to d^{-1/3} \nu_e E^{+1/2}_1,\;\; \ldots 
\eea 
These decays are superweak, leading to long-lived exotic charged leptons and quarks \cite{Frampton:2015cia}. 
More details on this topic can be found 
in \bib{Fairbairn:2006gg}. Experimental searches for long-lived charged particles similar to those have been carried on at the LHC \cite{Chatrchyan:2013oca}.   
From collider searches we can obtain lower limits on the masses. 

On the other hand, there must be constraints from cosmology. 
Qualitatively, the relic density ($\rho_X$) of the stable exotic-charged particles (named $X$) is proportional to 
the inverse of the annihilation cross section $\sigma_\text{ann}(X\bar{X}\to \text{SM})$, 
where SM here means a set of SM particles (see e.g. \bib{Kolb:1990vq}). 
Under this assumption, the model seems to be in conflict with the fact that 
none of $X$s (or its effects) has been to date noticed, because we may naively expect that 
$\sigma_\text{ann}$ cannot be too large compared to $\sigma_\text{EW}(\text{SM}\to \text{SM})$. 
This may be true if $X$ is a fermion and hence the above scenario of a stable fermion may be 
disfavored. However, if $X$ is a boson, e.g. $H^{\pm 1/2}$, then 
$\sigma_\text{ann}$ can be very large if e.g. $H_3$ is very heavy (i.e. $u\gg v$). A dominant 
mechanism can be $H^{+1/2} H^{-1/2} \to H_3 \to W^+ W^-$, whose amplitude is proportional 
to $M_{H_3}^2$. Of course, the cross section cannot exceed the unitary limit, but this kind of 
situation is a proof that $\sigma_\text{ann}$ can be much larger than $\sigma_\text{EW}$, leading to a 
smaller density. It may be interesting to perform a quantitative analysis to see how small the 
theoretical density can be. 
In this scenario of a stable boson, the exotic quarks are heavier. It is therefore more difficult to produce them 
at the LHC.  
Experimental searches for stable exotic-charged particles coming from outside the Earth can be found in 
Refs.~\cite{Perl:2001xi,Perl:2009zz,Agashe:2014kda}, where limits on the flux are given. 

\section{Conclusions}
\label{conclusions}
There are many 3-3-1 models. One important parameter to characterize the model is $\beta$. Which value of $\beta$ is the best 
to fit experimental data ? 
This question is to date still open. 
In this letter, we have discussed the special case of vanishing $\beta$. We showed that imposing the 
approximate global custodial symmetry at the SM energy scale on the scalar potential leads to 
a simple picture of the Higgs and gauge sectors at tree level. 
The $Z-Z'$ mixing vanishes at tree level and the value of the $\rho$ parameter can be kept close to one.  
Tree-level FCNCs are also reduced to three particles, namely the $Z'$, the CP-odd Higgs $A$ and the CP-even Higgs $H_3$. 

An important consequence of this consideration is the prediction of new leptons with 
electric charges of $\pm 1/2e$ and new quarks with $\pm 1/6e$ charges 
as well as new gauge and scalar bosons with $\pm 1/2e$ charges. 
Electric charge conservation requires that one of them must be stable. 
Their masses are unknown and they have never been 
experimentally observed. We think that they should be taken into account in experimental searches, whose results (if model-independent enough) will 
help to confirm or exclude many theoretical scenarios. If an experimental signature arises, the present model provides a very simple framework.
  
\section*{Acknowledgments}
We thank Hoang Ngoc Long for reading the manuscript and encouragement. 
LDN is grateful to Bas Tausk for fruitful discussions. 
LTH is funded by Vietnam  National
Foundation for Science and Technology Development (NAFOSTED) under
grant number 103.01-2014.69. 
The work of LDN has been partly supported by the German Ministry of Education and Research (BMBF) under contract
no. 05H15KHCAA. 


\end{document}